\documentstyle[sprocl,psfig]{article}
\def\NPB{{\em Nucl. Phys.} B}
\def\PLB{{\em Phys. Lett.}  B}
\def\PRL{\em Phys. Rev. Lett.}
\def\PRD{{\em Phys. Rev.} D}
\def\ZPC{{\em Z. Phys.} C}

\def\be{\begin{equation}}
\def\ee{\end{equation}}
\def\bea{\begin{eqnarray}}
\def\eea{\end{eqnarray}}
\def\be{\begin{equation}}
\def\ee{\end{equation}}
\def\bea{\begin{eqnarray}}
\def\eea{\end{eqnarray}}
\def\simlt{\stackrel{<}{{}_\sim}}
\def\simgt{\stackrel{>}{{}_\sim}}

\def\NPB#1#2#3{{\it Nucl.~Phys.} {\bf{B#1}} (19#2) #3}
\def\PLB#1#2#3{{\it Phys.~Lett.} {\bf{B#1}} (19#2) #3}
\def\PRD#1#2#3{{\it Phys.~Rev.} {\bf{D#1}} (19#2) #3}
\def\PRL#1#2#3{{\it Phys.~Rev.~Lett.} {\bf{#1}} (19#2) #3}
\def\ZPC#1#2#3{{\it Z.~Phys.} {\bf C#1} (19#2) #3}
\def\PTP#1#2#3{{\it Prog.~Theor.~Phys.} {\bf#1}  (19#2) #3}

\def\PR#1#2#3{{\it Phys.~Rep.} {\bf#1} (19#2) #3}
\def\RMP#1#2#3{{\it Rev.~Mod.~Phys.} {\bf#1} (19#2) #3}
\def\HPA#1#2#3{{\it Helv.~Phys.~Acta} {\bf#1} (19#2) #3}

\begin{document}

\begin{flushright}
IEM-FT-141/96 \\
 hep--ph/9609392 \\
\end{flushright}

\vspace{1cm}

\title{HIGGS BOSONS IN THE STANDARD MODEL AND THE\\
MINIMAL SUPERSYMMETRIC STANDARD MODEL 
\footnote{Based on lectures given at the {\it XXIV INTERNATIONAL MEETING
ON FUNDAMENTAL PHYSICS: From Tevatron to LHC}, 
22-26 April, 1996, Gand\'{\i}a (Valencia) Spain. }
 } 

\author{ M. QUIROS}

\address{Instituto de Estructura de la Materia, Serrano 123,\\
28006-Madrid, Spain}


\maketitle\abstracts{
\begin{center}
{\bf Abstract}
\end{center}
In these lectures we present a brief review of the Higgs
boson sector in the {\it Standard Model}, and its Minimal Supersymmetric
Extension, with particular emphasis on the main mechanisms for
Higgs production and decay at LEP2 and LHC, and theoretical bounds
on the Higgs boson masses. In the {\bf Standard Model} the effective 
potential can develop a non-standard minimum for values of
the field much larger than the weak scale. Comparison of
the decay rate to the non-standard minimum at finite (and
zero) temperature with the corresponding expansion rate 
of the Universe allows to identify the region, in the ($M_H$,
$M_t$)-plane which can be accomodated by the theory. 
In the {\bf Minimal Supersymmetric Standard Model},
approximate analytical expressions for the Higgs mass spectrum
and couplings are worked out. An appropriate
treatment of squark decoupling allows to consider large values of
the stop mixing parameters and thus fix a reliable upper
bound on the mass of the lightest CP-even Higgs boson mass.
The discovery of the Higgs boson at LEP2 might
put an upper bound (below the Planck scale) 
on the scale of new physics $\Lambda$ and eventually disentangle between
the Standard Model and the Minimal Supersymmetric Standard Model.}

\vspace{1cm}
\begin{flushleft}
IEM-FT-141/95\\
September 1995 \\
\end{flushleft}

\newpage

\section{Higgs bosons in the Standard Model}

In this lecture we will review some elementary and/or well established
features of the Higgs sector in the Standard Model (SM)~\cite{SM}. 
Most of it should
be viewed as an introduction for beginners and/or students
in the field though we also have presented some recent results on 
Higgs mass bounds obtained by this author in various collaborations. 
The methods used to obtain the latter results are sometimes technical.
Therefore, we have simplified the analysis and presented only the relevant
results.

\subsection{\bf Why a Higgs boson?}

The Higgs mechanism~\cite{Higgs} 
is the simplest mechanism to induce spontaneous
symmetry breaking of a gauge theory. In particular, in the Standard
Model of electroweak interactions it achieves the breaking
\begin{equation}
SU(2)_L \times U(1)_Y\longrightarrow U(1)_{em}
\label{rotura}
\end{equation}
in a renormalizable quantum field theory, and gives masses to the 
gauge bosons $W^{\pm},Z$, the Higgs boson and the fermions.

The SM fermions are given by~\cite{Pepe}
\begin{eqnarray}
\label{fermiones}
q_L & = & \left(
\begin{array}{c}
u_L \\
d_L
\end{array}
\right)_{1/6},\
\left(u_R\right)_{2/3},\
\left(d_R\right)_{-1/3} \nonumber\\
\ell_L & = & \left(
\begin{array}{c}
\nu_L \\
\ell_L
\end{array}
\right)_{-1/2},\
\left(\ell_R\right)_{-1}
\end{eqnarray}
where the hypercharge $Y$ is related to the electric charge $Q$ by,
$Q=T_3+Y$, and we are using the notation $f=f_L+f_R$, with
\begin{eqnarray}
\label{fermquirales}
f_L &= & \frac{1}{2}(1-\gamma_5)f \nonumber \\
f_R &= & \frac{1}{2}(1+\gamma_5)f .
\end{eqnarray}
The Higgs boson is an $SU(2)_L$ doublet, as given by
\begin{equation}
\label{higgsSM}
H=\frac{1}{\sqrt{2}}\left(
\begin{array}{c}
\chi^+ \\
\Phi+i\chi^0
\end{array}
\right)_{1/2}
\end{equation}
The physical Higgs $\phi$ is related to $\Phi$ by, $\Phi=\phi+v$, where
$v=(\sqrt{2}G_F)^{-1/2}=246.22$ GeV 
is the vacuum expectation value (VEV) of the Higgs.
The (massless) fields $\chi^\pm , \chi^0$ are the Goldstone
bosons.

A mass term for gauge bosons $V_{\mu}$, as
$\frac{1}{2}M_V^2 V_\mu V^\mu$
is not gauge invariant, and would spoil the renormalizability properties
of the theory. A mass term for fermions,
$m_u \bar{q}_L u_R+m_d \bar{q}_L d_R+m_\ell \bar{\ell}_L \ell_R$
does not even exist (it is not $SU(2)_L\times U(1)_Y$ invariant). Both goals
can be achieved through the Higgs mechanism~\cite{Higgs}.

One can  write the part of the SM Lagrangian giving rise to 
mass terms as
\begin{equation}
\label{lagrangiano}
{\cal L}=\left(D_\mu H\right)^\dagger \left(D_\mu H\right)
- (h_d \bar{q}_L H d_R + h_u \bar{q}_L H^c u_R+
h_\ell \bar{\ell}_LH\ell_R +h.c.) -V(H)
\end{equation}
where $H^c\equiv  i \sigma_2 H^*$, the covariant derivative
$D_\mu$ of the Higgs field is defined by
\begin{equation}
\label{dercovariante}
D_\mu H\equiv \left(\partial_\mu+i g \frac{\vec{\sigma}}{2} \vec{W}_\mu
+i g' \frac{1}{2} B_\mu \right) H
\end{equation}
and the Higgs potential by
\begin{equation}
\label{Higgspot}
V(H)=-\mu^2 H^\dagger H +\frac{\lambda}{2}\left(H^\dagger H\right)^2
\end{equation}
Minimization of (\ref{Higgspot}) yields,
\begin{equation}
\label{minimo}
\langle 0|H|0 \rangle\equiv \frac{v}{\sqrt{2}}
\left(
\begin{array}{c}
0 \\
1
\end{array}
\right);\ v=\sqrt{\frac{2\mu^2}{\lambda}}
\end{equation}

Replacing now $\Phi=\phi+v$ into (\ref{lagrangiano}) yields:
\begin{eqnarray}
\label{lagmasas}
{\cal L}& = & -\frac{1}{4}g^2v^2 W_\mu^+ W^{\mu -}
-\frac{1}{8}v^2
\left(
\begin{array}{cc}
Z^\mu & A^\mu 
\end{array}
\right)
\left(
\begin{array}{cc}
g^2+g'^2 & 0 \\
0  &  0
\end{array}
\right)
\left(
\begin{array}{c}
Z_\mu \\
A_\mu
\end{array}
\right) \nonumber \\
&&-\frac{vh_u}{\sqrt{2}}\bar{u}u
-\frac{vh_d}{\sqrt{2}}\bar{d}d
-\frac{vh_\ell}{\sqrt{2}}\bar{\ell}\ell
\end{eqnarray}
where
\begin{eqnarray}
W_\mu^{\pm}& = & \frac{1}{\sqrt{2}}
\left(W_{\mu 1}\pm i W_{\mu 2}\right) \nonumber\\
Z_\mu & = & \cos\theta_W W_{\mu 3}-\sin\theta_W B_\mu  \\
A_\mu & = & \sin\theta_W W_{\mu 3}+\cos\theta_W B_\mu \nonumber
\end{eqnarray}
and the electroweak angle $\theta_W$ is defined by 
$\tan\theta_W=g'/g$.

In this way the goal of giving masses to the gauge bosons and  the
fermions has then been achieved as \footnote{In the following we will 
use the notation $m_t,m_H$ for the top-quark and Higgs boson running
$\overline{\rm MS}$ on-shell masses
(defined at a scale equal to the corresponding mass), and $M_t,M_H$ for
the corresponding pole (physical) masses. They are related by a
contribution from self-energies. Thus for the Higgs boson, the running
and pole masses are related by~\cite{CEQR}
$M_H^2=m_H^2(M_H)+{\rm Re}\Pi_{\phi\phi}(M_H)-{\rm Re}\Pi_{\phi\phi}(0)$.}
\begin{eqnarray}
M_W^2 & = & \frac{1}{4} g^2 v^2 \nonumber \\
M_Z^2 & = & \frac{1}{4} \left(g^2+g'^2\right) v^2\\
m_f & = & \frac{1}{\sqrt{2}} h_f v \nonumber\\
m_H^2 & = & \lambda v^2\nonumber
\end{eqnarray}

\subsection{\bf What we know about the Higgs: its couplings}
 
The couplings $(g,g',v)$ are experimentally {\it traded} by
a set of three observables, as e.g. $(M_W,M_Z,G_F)$, or 
$(\alpha_{em},M_Z,G_F)$, while the Yukawa couplings $h_f$ are
{\it measured} by the fermion masses, $m_f$. Only the quartic
coupling $\lambda$ in Eq.~(\ref{lagrangiano}), which should
be {\it measured} by the Higgs mass, is 
at present {\bf unknown}.

All Higgs interactions (cross-sections, branching ratios,...)
are determined once the corresponding Feynman rules 
are known~\cite{Japs}.
In Table~1 we summarize the main vertices involving the physical
Higgs boson in the SM along with the rest of particles in the 
SM.

\begin{center}
\begin{tabular}{||c|c||}\hline
Vertex & Coupling \\ \hline
& \\
$\phi f\bar{f}$ & $-i\frac{g}{2M_W}m_f$ \\
& \\
$\phi W^{\pm}_\mu W^{\mp}_\nu$ & $i g M_W g_{\mu\nu}$ \\
&\\
$\phi Z_\mu Z_\nu$ & $ i\frac{g M_Z}{\cos\theta_W}g_{\mu\nu}$ \\
&\\
$\phi\phi\phi$ &$ -i\frac{3g}{2M_W}M^2_H $ \\
&\\
$\phi\phi W^{\pm}_\mu W^{\mp}_\nu $ & $i\frac{1}{2}g^2 g_{\mu\nu}$ \\
&\\
$\phi\phi Z_\mu Z_\nu$ & $i\frac{1}{2}\frac{g^2}{\cos^2\theta_W}g_{\mu\nu}$\\
&\\
$\phi\phi\phi\phi $ & $ -i\frac{3g^2M_H^2}{4M_W^2}$ \\
&\\
\hline
\end{tabular}
\end{center}

\begin{center}
Table 1
\end{center}

\subsubsection{Higgs production at LEP2}

\begin{figure}[hbt]
\centerline{
\psfig{figure=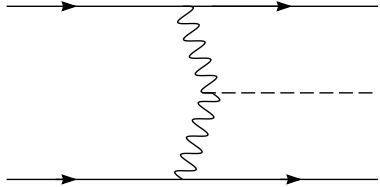,width=5cm,bbllx=4.75cm,bblly=11.5cm,bburx=14.25cm,bbury=17.cm}}
\caption{Vector-Vector fusion process for Higgs production.}
\label{diag2}
\end{figure}

The main mechanisms for production of Higgs particles at $e^+e^-$
colliders, at the LEP2 energies, are~\cite{LEP2}:
\begin{itemize}
\item
HIGGS-STRAHLUNG: $e^+e^-\rightarrow Z\phi$, where the Higgs boson
is radiated off the virtual $Z$-boson line exchanged in the s-channel.
[Fig.~\ref{diag1}, where the solid (fermion) lines are
electrons, the wavy line is a $Z$ boson and the dashed line a 
Higgs $\phi$.]
\begin{figure}[hbt]
\centerline{
\psfig{figure=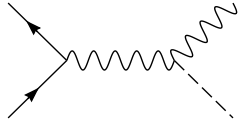,width=5cm,bbllx=4.75cm,bblly=14.5cm,bburx=14.25cm,bbury=17.7cm}}
\caption{Higgs-strahlung process for Higgs production.}
\label{diag1}
\end{figure}
\item
WW-FUSION: $e^+e^-\rightarrow \phi\bar{\nu}_e\nu_e$, where the Higgs
boson is formed in the fusion of virtual $WW$
exchanged in the t-channel. The virtual $W$'s 
are radiated off the electron and positron of the beam.
[Fig.~\ref{diag2}, where the incoming lower (upper) 
fermion line is an electron
(positron) and the corresponding outcoming fermion a $\nu_e$
($\bar{\nu}_e$). Wavy lines are $W$ and the dashed line a
Higgs.] 
\end{itemize}

A detailed analysis of these processes for LEP2 can be found in
Ref.~\cite{LEP2}. There it is found that the Higgs-strahlung process 
dominates the cross-section for low values of the Higgs mass
($M_H<105$ GeV), while the WW-fusion process dominates it for large 
values of the Higgs mass ($M_H>105$ GeV). 

\subsubsection{Higgs production at LHC}

The main mechanisms for production of Higgs bosons at $pp$ colliders,
at the LHC energies, are~\cite{Ferrando}:

\begin{figure}[hbt]
\centerline{
\psfig{figure=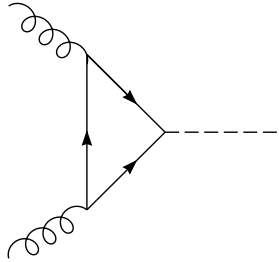,width=5cm,bbllx=4.75cm,bblly=12.7cm,bburx=14.25cm,bbury=19.2cm}}
\caption{Gluon-gluon fusion process for Higgs production.}
\label{diag4}
\end{figure}

\begin{figure}[hbt]
\centerline{
\psfig{figure=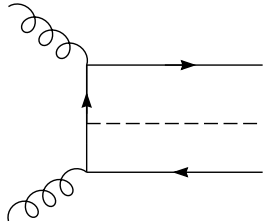,width=5cm,bbllx=4.75cm,bblly=13.cm,bburx=14.25cm,bbury=19.cm}}
\caption{Associated production of Higgs with $f\bar{f}$.}
\label{diag3}
\end{figure}

\begin{itemize}
\item
GLUON-GLUON FUSION: $gg\rightarrow \phi$, where two gluons in the sea
of the  protons collide through a loop of top-quarks, which
subsequently emits a Higgs boson. [Fig.~\ref{diag4} where the
curly lines are gluons, the internal fermion line a top and the
dashed line a Higgs.]
\item
WW (ZZ)-FUSION: $W^\pm W^\mp (ZZ)\rightarrow \phi$, where the Higgs
boson is formed in the fusion of $WW (ZZ)$, the virtual $W (Z)$'s 
being exchanged in the t-channel and radiated off 
a quark in the proton beam. [Fig.~\ref{diag2}, where wavy lines
are $W$ ($Z$), the incoming fermions quarks $q$ and the
outcoming fermions quarks $q$ ($q'$). The dashed line is the
Higgs.]
\item
HIGGS STRAHLUNG: $q \bar{q}^{(')}\rightarrow Z (W) \phi$,
where the Higgs boson is radiated off the virtual $Z(W)$-boson line
exchanged in the s-channel. [Fig.~\ref{diag1}, where wavy lines
are $Z$ ($W$), the incoming fermion a quark $q$ and the
outcoming fermion a quark $q$ ($q'$).]
\item
ASSOCIATED PRODUCTION WITH $t\bar{t}$: $gg\rightarrow \phi t \bar{t}$,
where the gluons from the proton sea exchange a top quark in the
t-channel, which emits a Higgs boson. [Fig.~\ref{diag3}, where
curly lines are gluons and the fermion line corresponds to a
quark $t$. The dashed line is the Higgs boson.]
\end{itemize}

A complete analysis of the different production channels can be found,
e.g. in Ref.~\cite{LHC}. It is found that for a top mass in the
experimental range~\cite{top} the gluon-gluon fusion mechanism is
dominating the production cross-section for any value of the Higgs
mass. The subdominant process, WW(ZZ)-fusion is comparable in magnitude
to the gluon-gluon process only for very large values of the Higgs
mass $M_H\sim 1$ TeV. For low values of the Higgs mass, $M_H\sim 100$ GeV,
the gluon-gluon fusion process is still dominant over all other channels
by around one order of magnitude, while all the others are similar
in magnitude for these values of the Higgs mass.

\subsubsection{Higgs decays}

For values of the Higgs mass relevant at LEP2 energies, the main
decay modes of the Higgs boson are:
\begin{itemize}
\item
$\phi\rightarrow b\bar{b},c\bar{c},\tau^-\tau^+$, which is dominated by the
$b\bar{b}$ channel.
\item
$\phi\rightarrow gg$, where the gluons are produced by a top-quark loop
emitted by the Higgs. [The inverse diagram of Fig.~\ref{diag4}.]
\item
$\phi\rightarrow WW^*\rightarrow W f \bar{f}'$, which is relevant for values
of the Higgs mass, $M_H>M_W$.
\end{itemize}

A complete analysis of different Higgs decay channels reveals~\cite{LEP2}
that, for LEP2 range of Higgs masses, $M_H<110$ GeV, the $b\bar{b}$
channel dominates the Higgs branching ratio by $\sim$ one order
of magnitude.

For $M_H>110$ GeV, the main decay modes relevant for LHC energies and
$pp$ colliders are~\cite{LHC}:
\begin{itemize}
\item
$\phi\rightarrow \gamma\gamma$, where the photons are produced by a top-quark
loop emitted by the Higgs. [The inverse diagram as that of
Fig.~\ref{diag4}, where gluons are replaced by photons.]
\item
$\phi\rightarrow W^{\pm}W^{\mp}$, which requires $M_H>2M_W$.
\item
$\phi\rightarrow ZZ$, which requires $M_H>2M_Z$.
\item
$\phi\rightarrow t\bar{t}$, which requires $M_H>2M_t$.
\end{itemize}

For a heavy Higgs ($M_H>150$ GeV) the $WW (ZZ)$ decay channels
completely dominate the Higgs branching ratio, while the radiative
decay $\gamma\gamma$ dominates for low values of the Higgs mass and is
expected to close the LHC window for a light Higgs. The reader
is referred to Ref.~\cite{LHC} for more details.

\subsection{\bf What we do not know about the Higgs: its mass}

Being the Higgs boson the missing ingredient of the SM, the quartic
coupling $\lambda$, and so its mass, are unknown. However we can
have information on $M_h$ from experimental and theoretical input.

From experimental inputs we have direct and indirect information
on the Higgs mass.
Since direct
experimental searches at LEP have been negative up to now,
they translate into a lower bound on the Higgs mass~\cite{Higgsmass}, 
\begin{equation}
\label{expbound}
M_h>67\ {\rm GeV},
\end{equation}
Experimental searches also yield
indirect information, which is the influence the Higgs mass
has in radiative corrections and in precision measurements at
LEP~\cite{Higgsmass}. 
However, unlike the top quark mass, on which the radiative
corrections are quadratically dependent, and so very
sensitive, the dependence of one-loop radiative corrections on
the Higgs mass is only logarithmic (the so-called Veltman's
screening theorem), which means that radiative corrections in
the SM have very little sensitivity to the Higgs mass, providing
only very loose bounds from precision measurements.

However, from the theoretical input the situation is rather
different. In fact the theory has a lot of information on $M_h$,
which can be used to put bounds on the Higgs mass. If these
bounds were evaded when the Higgs mass will be eventually
measured, this measurement might lead to the requirement of new physics,
just because the SM cannot accomodate such a value of the Higgs
(and the top-quark) mass.

For particular values of the Higgs boson and top quark masses, $M_H$ and $M_t$,
the effective potential of the Standard Model (SM) develops a deep non-standard
minimum for values of the field $\phi \gg G_F^{-1/2}$~\cite{L}. 
In that case the 
standard electroweak (EW) minimum becomes metastable and might decay into
the non-standard one. This means that the SM might have troubles 
in certain regions
of the plane ($M_H$,$M_t$), a fact 
which can be intrinsically interesting as evidence for
new physics. Of course, the mere existence of the non-standard minimum, 
and also the decay rate 
of the standard one into it, depends on the scale $\Lambda$ up to which
we believe the SM results. In fact, one can identify $\Lambda$ 
with the scale of new physics. 

\subsubsection{Stability bounds}

The preliminary question one should ask is: When the standard EW 
minimum becomes
metastable, due to the appearance of a deep non-standard 
minimum? This question was
addressed in past years~\cite{L} taking into account leading-log (LL) and 
part of next-to-leading-log (NTLL) corrections. 
More recently, calculations have incorporated all 
NTLL
corrections~\cite{AI,CEQ} 
resummed to all-loop by the renormalization group equations (RGE),
and considered pole masses for the top-quark and 
the Higgs-boson. 
From the requirement of a stable (not metastable) standard EW minimum 
we obtain a lower bound on
the Higgs mass, as a function of the top mass, labelled by the values of 
the SM cutoff (stability bounds). Our
result~\cite{CEQ} is lower than previous estimates by ${\cal O}$(10) GeV.
The problem to attack is easily stated as follows:

The effective potential in the SM can be written as (\ref{Higgspot})
\be
\label{poteff}
V=-\frac{1}{2}m^2\phi^2+\frac{1}{8}\lambda\phi^4+\cdots
\ee
where the ellipsis refers to radiative corrections and all
parameters and fields in (\ref{poteff}) are running with the
renormalization group scale $\mu(t)=M_Z\exp(t)$. The condition for 
having an extremal is 
$V'(\phi(t))=0$, which has as solution
\be
\label{vev}
\phi^2=\frac{2m^2}{\lambda-\frac{12}{32\pi^2}h_t^4
\left(\log\frac{h_t^2\phi^2}{2\mu^{2}}-1\right)}
\ee
where $h_t$ refers to the top Yukawa coupling, and only the
leading radiative corrections have been kept for simplicity. 
The curvature of the potential (\ref{poteff}) at the extreme 
is given by
\be
\label{curv}
V''(\phi)=2m^2+\frac{1}{2}\beta_\lambda \phi^2
\ee
The condition $V'=0$ is obviously satisfied at the EW minimum where
$\langle\phi\rangle=v\sim 246$ GeV, $\lambda\sim(m_H/v)^2>1/16$,
$m^2\sim m_H^2/2$ and $V''(\langle\phi\rangle)>0$ (a minimum).
However, the condition $V'=0$ can also be satisfied for values
of the field $\phi\gg v$ and, since $m={\cal O}(100)$ GeV, 
for those values
$$
\lambda\sim\left(\frac{m}{\phi}\right)^2\ll 1.
$$
Therefore, for the non-standard extremals we have 
\begin{eqnarray}
\label{minmax}
\beta_\lambda < 0 & \Longrightarrow & V''<0\ {\rm
maximum}\nonumber \\
\beta_\lambda > 0 & \Longrightarrow & V''>0\ {\rm minimum}. 
\end{eqnarray}

\begin{figure}[hbt]
\centerline{
\psfig{figure=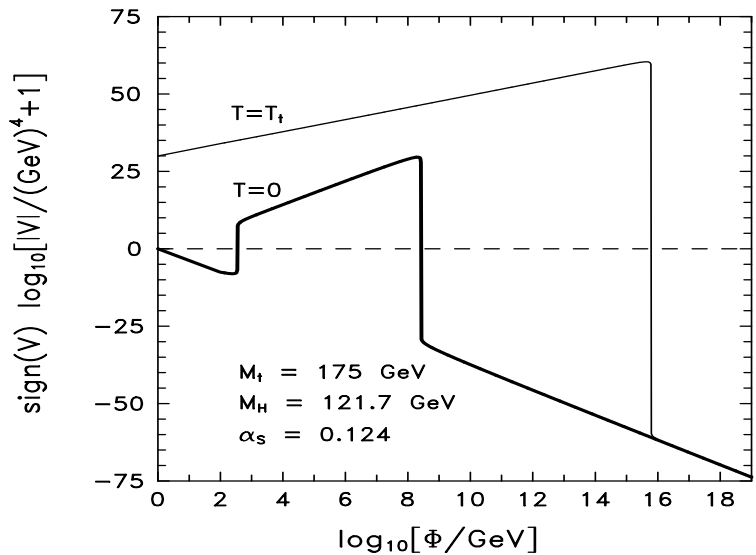,height=7.5cm,width=7cm,bbllx=4.75cm,bblly=3.cm,bburx=14.25cm,bbury=16cm}}
\caption{Plot of the effective potential for $M_t=175$ GeV, $M_H=121.7$
GeV at $T=0$ (thick solid line) and $T=T_t=2.5\times 10^{15}$ GeV 
(thin solid line).}
\label{fval1}
\end{figure}

The one-loop effective potential of the SM improved by 
two-loop RGE has been shown to
be highly scale independent~\cite{CEQR} and, therefore, very reliable for the 
present study. 
In Fig.~\ref{fval1} we show (thick solid line) 
the shape of the effective potential for 
$M_t=175$ GeV 
and $M_H=121.7$ GeV. We see the appearance of the non-standard maximum,
$\phi_M$, while the global
non-standard minimum has been cutoff at $M_{P\ell}$. 
We can see from Fig.~\ref{fval1} the 
steep descent from the non-standard maximum. Hence, 
even if the non-standard minimum is beyond 
the SM cutoff, the
standard minimum becomes metastable and might be destabilized. So for fixed 
values of $M_H$ and
$M_t$ the condition for the standard minimum not to become metastable is 
\be
\label{condstab}
\phi_M \simgt \Lambda
\ee
Condition (\ref{condstab}) makes the stability condition $\Lambda$-dependent. 
In fact we have plotted
in Fig.~\ref{fval2} the stability condition on $M_H$ versus $M_t$ for 
$\Lambda=
10^{19}$ GeV and 10 TeV. The stability 
region corresponds to the region above the dashed curves.
%
%
\subsubsection{Metastability bounds}

\begin{figure}[hbt]
\centerline{
\psfig{figure=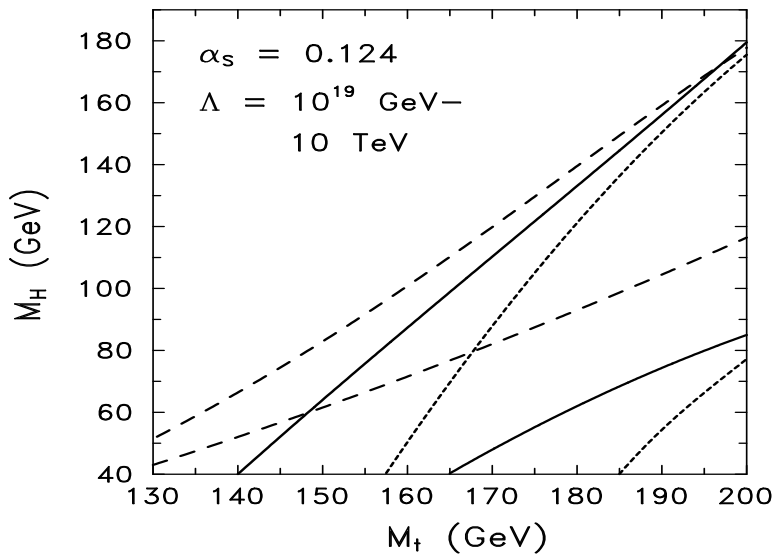,height=7.5cm,width=7cm,bbllx=5.cm,bblly=2.cm,bburx=14.5cm,bbury=15cm}}
\caption{Lower bounds on $M_H$ as a function of $M_t$, for
$\Lambda=10^{19}$ GeV (upper set) and $\Lambda=10$ TeV (lower set). 
The dashed curves
correspond to the stability bounds and  the solid (dotted) 
ones to the metastability bounds at finite (zero) temperature.}
\label{fval2}
\end{figure}

In the last subsection we have seen that 
in the region of Fig.~\ref{fval2}
below the dashed line the standard EW minimum is
metastable. However we should not draw physical consequences 
from this fact since we still do not
know at which minimum does the Higgs field sit. Thus, the real physical 
constraint we have to impose is avoiding
the Higgs field sitting at its non-standard minimum. 
In fact the Higgs field can be sitting at its 
zero temperature non-standard minimum because:
\begin{enumerate}
\item
The Higgs field was driven from the origin to the non-standard minimum 
at finite temperature 
by thermal fluctuations in a non-standard EW phase transition at 
high temperature. 
This minimum evolves naturally to the non-standard minimum at zero 
temperature. In this case 
the standard EW phase transition, at $T\sim 10^2$ GeV, will not take place.
\item
The Higgs field was driven from the origin to the 
standard minimum at $T\sim 10^2$ GeV, but decays,
at zero temperature, to the non-standard minimum by a quantum fluctuation.
\end{enumerate}
%
In Fig.~\ref{fval1} we have depicted the 
effective potential at $T=2.5\times 
10^{15}$ GeV (thin solid line) which
is the corresponding 
transition temperature. Our finite temperature 
potential~\cite{EQtemp} incorporates plasma effects~\cite{Q} 
by one-loop resummation of Debye masses~\cite{DJW}.  The tunnelling 
probability per unit time per
unit volume  was computed long ago for thermal~\cite{Linde} and 
quantum~\cite{Coleman} fluctuations.
At finite temperature it is given by $\Gamma/\nu\sim T^4 \exp(-S_3/T)$, 
where $S_3$ is the euclidean action evaluated
at the bounce solution $\phi_B(0)$. The semiclassical 
picture is that unstable bubbles are nucleated behind the
barrier at $\phi_B(0)$ with a probability given by $\Gamma/\nu$. Whether 
or not they fill the Universe depends on
the relation between the probability rate and the expansion 
rate of the Universe. By normalizing the former
with respect to the latter we obtain a normalized probability $P$, 
and the condition for decay corresponds
to $P\sim 1$. Of course our results are trustable,  
and the decay actually happens, only if
$\phi_B(0)<\Lambda$, so that the similar condition to (\ref{condstab}) is 
\be
\label{condmeta}
\Lambda< \phi_B(0)
\ee
The condition of no-decay (metastability condition) has 
been plotted in Fig.~\ref{fval2}  (solid lines)
for $\Lambda=10^{19}$ GeV and 10 TeV. The region
between the dashed and the solid line corresponds to 
a situation where the non-standard minimum exists
but there is no decay to it at finite temperature. 
In the region below the solid lines the Higgs field is sitting
already at the non-standard minimum at $T\sim 10^2$ GeV, and the  standard EW 
phase transition does not happen.

We also have evaluated the tunnelling probability 
at zero temperature from the standard EW minimum to the
non-standard one. The result of the calculation 
should translate, as in the previous case, in lower bounds
on the Higgs mass for differentes 
values of $\Lambda$. The corresponding bounds are shown 
in Fig.~\ref{fval2} in
dotted lines. Since the dotted lines 
lie always below the solid ones, the possibility of quantum tunnelling at
zero temperature does not impose any extra constraint.

As a consequence of all improvements in the 
calculation, our bounds are lower than previous 
estimates~\cite{AV}. To fix ideas, for $M_t=175$ GeV, the 
bound reduces by $\sim 10 $ GeV for $\Lambda=10^4$ GeV,
and $\sim 30$ GeV for $\Lambda=10^{19}$ GeV.

\subsubsection{Perturbativity bounds}

Up to here we have described lower bounds on the Higgs mass
based on stability arguments. Another kind of bounds, which have
been used in the literature, are upper bounds based on the
requirement of perturbativity of the SM up to the high scale
(the scale of new physics) $\Lambda$. 

\begin{figure}[hbt]
\centerline{
\psfig{figure=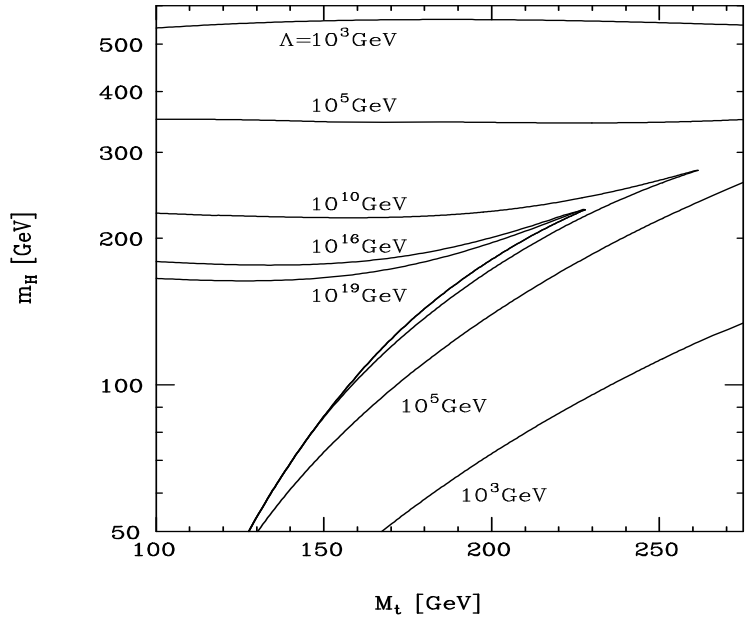,height=10cm,width=17cm,angle=90}}
\caption{Perturbativity and stability bounds on the SM Higgs
boson. $\Lambda$ denotes the energy scale where the particles
become strongly interacting.}
\label{lepp}
\end{figure}

Since the quartic coupling grows with the scale~\footnote{In fact
the value of the renormalization scale where the quartic
coupling starts growing depends on the value of the top-quark
mass.}, it will blow up to infinity at a given scale: the scale where
$\lambda$ has a Landau pole. The position of the Landau pole
$\Lambda$ is, by definition, the maximum scale up to which the
SM is perturbatively valid. In this way assuming the SM remains
valid up to a given scale $\Lambda$ amounts to requiring an
upper bound on the Higgs mass from the perturbativity 
condition~\cite{LEP2} 
\be
\label{perturbcond}
\frac{\lambda(\Lambda)}{4\pi}\leq 1
\ee
This upper bound depends on the scale $\Lambda$ and very mildly
on the top-quark mass $M_t$ through its influence on the
renormalization group equations of $\lambda$. We have plotted in
Fig.~\ref{lepp} this upper bound for different values of the high scale
$\Lambda$, along with the corresponding stability bounds.

\subsection{\bf A light Higgs can {\it measure} the scale of New Physics}

\begin{figure}[htb]
\centerline{
\psfig{figure=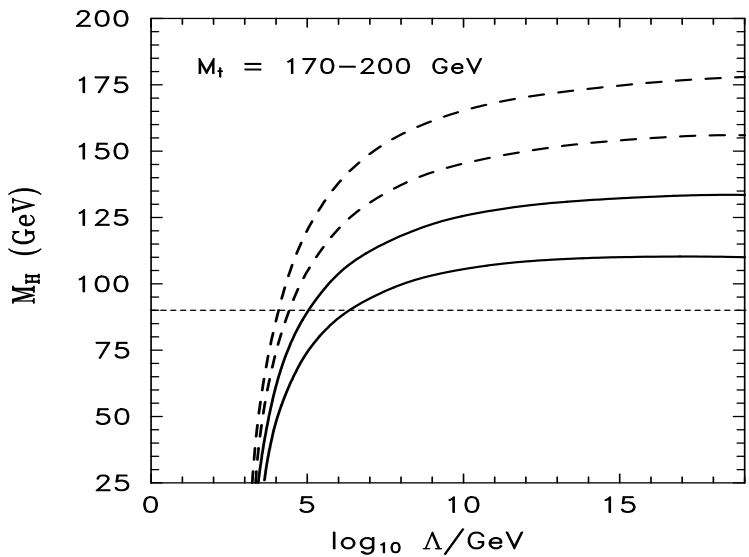,height=7.5cm,width=7cm,bbllx=5.cm,bblly=2.5cm,bburx=14.5cm,bbury=15.5cm}}
\caption{SM lower bounds on $M_H$ from
metastability requirements as a function of 
$\Lambda$ for different values of $M_t$.}
\label{fval6}
\end{figure}

From the bounds on  $M_H(\Lambda)$ previously obtained 
(see Fig.~\ref{fval6}) 
one can easily deduce that
a measurement of $M_H$ might provide an 
{\bf upper bound}  (below the Planck scale) on the
scale of new physics provided that 
\be
\label{final}
M_t>\frac{M_H}{2.25\;  {\rm GeV}}+123\; {\rm GeV}
\ee
Thus, the present 
experimental bound from LEP, $M_H>67$ GeV, would imply, from
(\ref{final}), $M_t>153$ GeV, which is fulfilled 
by experimental detection of the 
top~\cite{top}. Even non-observation of the Higgs at 
LEP2 (i.e. $M_H\simgt 95$ GeV), would
leave an open window ($M_t\simgt 165$ GeV) 
to the possibility that a future Higgs detection
at LHC could lead to an upper bound on $\Lambda$. Moreover, Higgs 
detection at
LEP2 would put an upper bound on the scale of new physics. Taking, 
for instance,  $M_H\simlt 95$
GeV and  170 GeV $< M_t< $ 180 GeV, then $\Lambda\simlt 10^7$ GeV, while for
180 GeV $< M_t <$ 190 GeV, $\Lambda\simlt 10^4$ 
GeV, as can be deduced from Fig.~\ref{fval6}. Finally, using as upper
bound for the top-quark mass $M_t<180$ GeV [Ref.~\cite{top}] we obtain
from (\ref{final}) that only if the condition
\be
M_h>128\ {\rm GeV}
\ee
is fulfilled, the SM can be a consistent theory up to the Planck
scale, where gravitational effects can no longer be neglected.

\section{Higgs bosons in the Minimal Supersymmetric Standard Model}

The Minimal Supersymmetric Standard Model 
(MSSM)~\cite{susy} is the best
motivated extension of the SM where some of their theoretical problems 
(e.g. the hierarchy problem inherent with the fact that the
SM cannot be considered as a fundamental theory for energies
beyond the Planck scale) find at least a technical solution~\cite{Carlos}. 
In this lecture we will concentrate on the Higgs
sector of the MSSM that is being the object of experimental
searches at present accelerators (LEP), and will equally be one of the
main goals at future colliders (LHC).

\subsection{\bf The Higgs sector in the Minimal Supersymmetric Standard 
Model}

The Higgs sector of the MSSM~\cite{Hunter} requires two Higgs doublets, with
opposite hypercharges, as
\begin{equation}
\label{higgsmssm}
H_1  =  \left(
\begin{array}{c}
H_1^0 \\
H_1^-
\end{array}
\right)_{-1/2}, \ \
H_2  =  \left(
\begin{array}{c}
H_2^+ \\
H_2^0
\end{array}
\right)_{1/2}
\end{equation}
The reason for this duplicity is twofold. On the one hand it is
necessary to cancel the triangular anomalies generated by the
higgsinos. On the other hand it is required by the structure of
the supersymmetric theory to give masses to all fermions.

The most general gauge invariant scalar potential is given,
for a general two-Higgs doublet model, by:
\begin{eqnarray}
\label{higgs2}
V& = & m_1^2 |H_1|^2+m_2^2|H_2|^2+(m_3^2 H_1 H_2+h.c.)
+\frac{1}{2}\lambda_1(H_1^\dagger H_1)^2\nonumber\\
&&+\frac{1}{2}\lambda_2(H_2^\dagger
H_2)^2+\lambda_3(H_1^\dagger H_1)(H_2^\dagger H_2)
+\lambda_4(H_1 H_2)(H_1^\dagger H_2^\dagger) \\
&&+\left\{\frac{1}{2}\lambda_5(H_1 H_2)^2+
\left[\lambda_6(H_1^\dagger H_1)+\lambda_7(H_1^\dagger
H_2^\dagger) \right](H_1 H_2)+h.c.\right\} \nonumber
\end{eqnarray}
However, supersymmetry provides the following tree-level
relations between the previous couplings. The non-vanishing ones are:
\begin{equation}
\label{lambdastree}
\lambda_1  = \lambda_2=\frac{1}{4}(g^2+g'^2), \
\lambda_3  = \frac{1}{4}(g^2-g'^2),\ 
\lambda_4  = -\frac{1}{4}g^2  
\end{equation}
Replacing (\ref{lambdastree}) into (\ref{higgs2}) one obtains
the tree-level potential of the MSSM, as:
\begin{eqnarray}
\label{potmssm}
V_{\rm MSSM}& = & m_1^2 H_1^\dagger H_1+m_2^2 H_2^\dagger H_2
+m_3^2(H_1 H_2+h.c.) \\
&&+\frac{1}{8}g^2\left(H_2^\dagger \vec{\sigma} H_2+
H_1^\dagger\vec{\sigma}H_1\right)^2
+\frac{1}{8}g'^2\left(H_2^\dagger H_2-H_1^\dagger H_1\right)^2
\nonumber
\end{eqnarray}

This potential, along with the gauge and Yukawa couplings in the
superpotential,
\be
\label{superp}
W=h_u Q\cdot H_2 U^c+h_d Q\cdot H_1 D^c+ h_\ell L\cdot H_1 E^c
+\mu H_1\cdot H_2
\ee
determine all couplings and masses (at the tree-level) of the
Higgs sector in the MSSM.

After gauge symmetry breaking,
\begin{eqnarray}
v_1 & = & \langle {\rm Re}\; H_1^0 \rangle \nonumber \\
v_2 & = & \langle {\rm Re}\; H_2^0 \rangle 
\end{eqnarray}
the Higgs spectrum contains one neutral CP-odd Higgs $A$
(with mass $m_A$, that will be taken as a free parameter)
\be
A=\cos\beta\;{\rm Im}H_2^0+\sin\beta\;{\rm Im}H_1^0
\ee
and one neutral Goldstone $\chi^0$
\be
\chi^0=-\sin\beta\;{\rm Im}H_2^0+\cos\beta\;{\rm Im}H_1^0
\ee
with $\tan\beta=v_2/v_1$. It also contains one complex charged
Higgs $H^\pm$,
\be
H^+=\cos\beta\; H_2^+ +\sin\beta\;(H_1^-)^*
\ee
with a (tree-level) mass
\be
\label{masapm}
m_{H^\pm}^2=M_W^2+m_A^2
\ee
and one charged Goldstone $\chi^\pm$,
\be
\chi^+=-\sin\beta\; H_2^+ +\cos\beta\;(H_1^-)^*.
\ee
Finally the Higgs spectrum contains two CP-even neutral Higgs
bosons $H,{\cal H}$ (the light and the heavy mass eigenstates)
which are linear combinations of Re~$H_1^0$ and
Re~$H_2^0$, with a mixing angle $\alpha$ given by
\be
\label{mixingHiggs}
\cos 2\alpha=-\cos2\beta\;\frac{m_A^2-M_Z^2}{m_{\cal H}^2-m_H^2}
\ee
and masses
\be
\label{masahH}
m^2_{H,{\cal H}}=\frac{1}{2}\left[
m_A^2+M_Z^2\mp\sqrt{(m_A^2+M_Z^2)^2-4m_A^2M_Z^2\cos^2 2\beta}
\right]
\ee

\subsubsection{The Higgs couplings}

All couplings in the Higgs sector are functions of the gauge
($G_F,g,g'$) and Yukawa couplings, as in the SM, and of the
previously defined mixing angles $\beta,\alpha$. 
Some relevant couplings are contained in Table~2
where all particle momenta, in squared brackets,  are incoming.

\begin{center}
\begin{tabular}{||c|c||}\hline
Vertex & Couplings \\ \hline
& \\
$(H,{\cal H})WW $ & $(\phi WW)_{\rm
SM}[\sin(\beta-\alpha),\cos(\beta-\alpha)]$ \\
& \\
$(H,{\cal H})ZZ $ & $(\phi ZZ)_{\rm
SM}[\sin(\beta-\alpha),\cos(\beta-\alpha)]$ \\
& \\
$(H,{\cal H},A)[p]W^\pm H^\mp [k] $ & $\mp i\frac{g}{2}(p+k)^\mu
[\cos(\beta-\alpha), -\sin(\beta-\alpha),\pm i]$ \\
& \\
$(H,{\cal H},A)u\bar{u} $  & $(\phi u\bar{u})_{\rm
SM}[{\displaystyle \frac{\cos\alpha}{\sin\beta},
\frac{\sin\alpha}{\sin\beta} , -i\gamma_5 \cot\beta]} $  \\
& \\
$(H,{\cal H},A)d\bar{d} $  &$(\phi d\bar{d})_{\rm
SM}[{\displaystyle -\frac{\sin\alpha}{\cos\beta},
\frac{\cos\alpha}{\cos\beta}, -i\gamma_5 \tan\beta] } $ \\
& \\
$H^- u\bar{d} $
& $ {\displaystyle \frac{ig}{2\sqrt{2}M_W}
[(m_d\tan\beta+m_u\cot\beta) - (m_d
\tan\beta -m_u \cot\beta)\gamma_5] } $ \\
& \\
$ H^+ \bar{u} d $ & $ {\displaystyle \frac{ig}{2\sqrt{2}M_W}
[(m_d\tan\beta+m_u\cot\beta) + (m_d
\tan\beta -m_u \cot\beta)\gamma_5] } $ \\
& \\
$(\gamma,Z)H^+[p]H^- [k] $ & $ {\displaystyle 
-i(p+k)^\mu\left[e,g\frac{\cos
2\theta_W}{2 \cos\theta_W}\right] } $ \\
& \\
$h[p] A [k] Z $ & ${\displaystyle 
-\frac{e}{2\cos\theta_W\sin\theta_W} (p+k)^\mu
\cos(\beta-\alpha) } $ \\
& \\ \hline
\end{tabular}
\end{center}

\vspace{1cm}
\begin{center}
Table 2
\end{center}

\subsubsection{Higgs production at LEP2}

The main mechanisms for production of 
neutral Higgs particles at $e^+e^-$
colliders, at the LEP2 energies, are~\cite{LEP2}:

\begin{itemize}
\item
HIGGS-STRAHLUNG: $e^+e^-\rightarrow ZH$, where the Higgs boson
is radiated off the virtual $Z$-boson line. This process is
identical to the SM Higgs-strahlung. [See Fig.~\ref{diag1}.]
\item
ASSOCIATED PAIR PRODUCTION: $e^+ e^- \rightarrow HA$,
$e^+ e^-\rightarrow H^\pm H^\mp$. 
The production of $HA$ is mediated by a $Z$-boson in the
s-channel (it uses the coupling hAZ in Table~2). The
production of $H^\pm H^\mp$ can be mediated by either $\gamma$
and $Z$, using the $(\gamma,Z)H^\pm H^\mp$ vertex in Table~2. 
\end{itemize}
A detailed analysis of these processes for LEP2 can be found in
Ref.~\cite{LEP2}.

\subsubsection{Higgs production at LHC}

The main mechanisms for production of 
neutral Higgs bosons at $pp$ colliders,
at the LHC energies, are~\cite{Ferrando}:

\begin{itemize}
\item
GLUON-GLUON FUSION: $gg\rightarrow (H,{\cal H},A)$, 
where two gluons in the sea
of the protons collide through a loop of top-quarks, bottom-quarks,
stops and sbottoms which
subsequently emit a Higgs boson. The contribution of a (s)bottom
loop is only relevant for large values of $\tan\beta$.
[Figs.~\ref{diag4} and \ref{diag5}, where curly lines are
gluons, internal fermion lines quarks $t$ and $b$, internal
boson (dashed) lines squarks $\tilde{t}$ and $\tilde{b}$ and the
dashed line is a Higgs boson $H$, ${\cal H}$ or $A$.] 
\begin{figure}[hbt]
\centerline{
\psfig{figure=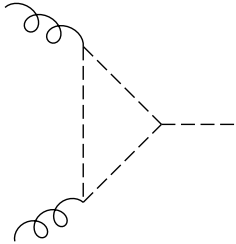,width=5cm,bbllx=4.75cm,bblly=10.5cm,bburx=14.25cm,bbury=16.5cm}}
\caption{Gluon-gluon fusion process for Higgs production with a squark loop.}
\label{diag5}
\end{figure}

\item 
WW (ZZ)-FUSION: $W^\pm W^\mp \rightarrow (H,{\cal H},A)$, 
$ZZ \rightarrow (H,{\cal H},A)$, where the Higgs
boson is formed in the fusion of $WW (ZZ)$, the virtual $W (Z)$'s 
being radiated off a quark in the proton beam. [See
Fig.~\ref{diag2} where the external dashed line corresponds to a
Higgs boson $H$, ${\cal H}$ or $A$.]
\item
HIGGS STRAHLUNG: $q \bar{q}\rightarrow Z  (H,{\cal H},A)$,
$q \bar{q}'\rightarrow W (H,{\cal H},A)$
where the corresponding Higgs boson
is radiated off the virtual $Z(W)$-boson line. [See
Fig.~\ref{diag1}, where the dashed line is a Higgs boson, $H$,
${\cal H}$  or $A$.]  
\item
ASSOCIATED PRODUCTION WITH $t\bar{t},b\bar{b}$: 
$gg\rightarrow  t \bar{t} (H,{\cal H},A)$,
$gg\rightarrow b \bar{b} (H,{\cal H},A)$
where the gluons from the proton sea exchange a top 
(bottom)-quark in the
t-channel, the exchanged top (bottom) quark emitting a Higgs boson.
[See Fig.~\ref{diag3} where the curly lines are gluons, the
fermion line a $t$ or $b$ quark and the dahsed line a Higgs
boson $H$, ${\cal H}$ or $A$.]
\end{itemize}

\begin{figure}[hbt]
\centerline{
\psfig{figure=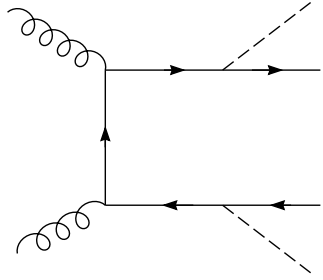,width=5cm,bbllx=5.75cm,bblly=12.cm,bburx=15.25cm,bbury=19.5cm}}
\caption{Charged higgs production process.}
\label{diag6}
\end{figure}

The production of a charged Higgs boson is through the process
$gg\rightarrow t\bar{t}$, where the gluons exchange a top-quark
in the t-channel, and subsequent decay $t\rightarrow b H^+$.
This process is available only when $M_t>m_{H^+}+M_b$. Otherwise
the detection of the charged Higgs is much more difficult.
[Fig.~\ref{diag6} where curly lines are gluons, the fermion
exchanged between the gluons a $t$ quark, the external fermions
$b$ quarks and the external bosons (dashed) are $H^\pm$.]
A complete analysis of the different production channels can be found,
e.g. in Ref.~\cite{FabioLHC}. 

\subsubsection{Higgs decays}

Assuming R-parity conservation, two-body decays should be into
SM particles, or two supersymmetric partners if the
supersymmetric spectrum is kinematically accesible. Assuming the
supersymmetric spectrum to be heavy enough (a useful working
hypothesis), the decays are always into SM particles. 
The main decay modes of the Higgs boson are then:
\begin{itemize}
\item
$(H,{\cal H},A)\rightarrow b\bar{b},c\bar{c},\tau^-\tau^+,t\bar{t},gg,
\gamma\gamma,W^* W^*, Z^* Z^*,Z\gamma$, which is very similar
to the corresponding SM modes.
\item
$H\rightarrow AA$.
\item
${\cal H}\rightarrow hh,AA,ZA$.
\item
$A\rightarrow ZH$:
\item
$H^+\rightarrow c\bar{s},\tau^+\nu_\tau,t\bar{b},W^+ H$.
\end{itemize}
A complete analysis of the decay modes in the MSSM can be found in
Ref.~\cite{LEP2}, for LEP2, and~\cite{FabioLHC} for LHC.

\subsection{\bf Radiative corrections}

All previous Higgs production and decay processes depend on the
Higgs masses $m_H,m_{\cal H},m_A,m_{H^\pm}$, and couplings
$g,g',G_F,\tan\beta,\cos\alpha,h_f,\lambda_1,\dots,\lambda_7$. 
We have already given their tree-level values. In particular,
the mass spectrum satisfies at tree-level the following
relations: 
\begin{eqnarray}
\label{treerel}
m_H & < & M_Z|\cos 2\beta| \nonumber \\
m_H & < & m_A \\
m_{H^\pm} & > & M_W \nonumber
\end{eqnarray}
which could have a number of very important phenomenological
implications, as it is rather obvious. However, it was
discovered that radiative corrections are important and can
spoil the above tree level relations with a great
phenomenological relevance. A detailed knowledge of radiatively
corrected couplings and masses is necessary for experimental
searches in the MSSM.

The {\bf effective potential} methods to compute the (radiatively
corrected) Higgs mass spectrum in the 
MSSM are useful since they allow to {\bf resum}
(using Renormalization Group (RG) techniques) LL,
NTLL,..., corrections to {\bf all orders}
in perturbation theory. These methods~\cite{Effpot,EQ}, as well as the
diagrammatic methods~\cite{Diagram} to compute the Higgs mass spectrum
in the MSSM, were first developed in the early nineties.

Effective potential methods are based on the {\bf run-and-match}
procedure by which all dimensionful and dimensionless couplings
are running with the RG scale, for scales greater than the
masses involved in the theory. When the RG scale
equals a particular mass threshold, heavy fields decouple,
eventually leaving threshold effects in order to match the
effective theory below and above the mass threshold. For
instance, assuming a common soft supersymmetry breaking mass 
for left-handed and right-handed stops and sbottoms, 
$M_S\sim m_Q\sim m_U\sim m_D$, and assuming for the top-quark mass, 
$m_t$, and for the CP-odd Higgs mass, $m_A$, the range 
$m_t\leq m_A\leq M_S$, we have: for scales $Q\geq M_S$, the MSSM, for
$m_A\leq Q\leq M_S$ the two-Higgs doublet model (2HDM), and for
$m_t\leq Q\leq m_A$ the SM. Of course there are
thresholds effects at $Q=M_S$ to match the MSSM with the 2HDM, and
at $Q=m_A$ to match the 2HDM with the SM. 
\begin{figure}[htb]
\centerline{
\psfig{figure=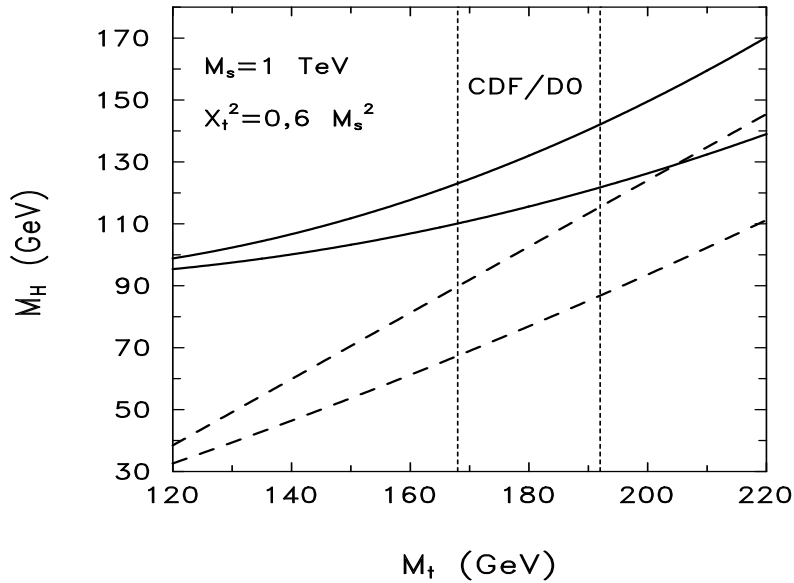,height=7.5cm,width=7cm,bbllx=5.5cm,bblly=2.5cm,bburx=15.cm,bbury=15.5cm}}
\caption{Plot of $M_H$ as a function of $M_t$ for $\tan\beta\gg 1$
(solid lines), $\tan\beta=1$ (dashed lines), and $X_t^2=6 M_S^2$ (upper set),
$X_t=0$ (lower set). The experimental band from the CDF/D0 detection is also
indicated.}
\label{fval3}
\end{figure}

As we have said, the neutral Higgs sector of the MSSM contains,  
on top of the CP-odd Higgs $A$, two CP-even Higgs mass
eigenstates, ${\cal H}$ (the heaviest one) and $H$ (the lightest one). 
It turns out that the larger
$m_A$ the heavier the lightest Higgs $H$. Therefore the case
$m_A\sim M_S$ is, not only a great simplification since the effective
theory below $M_S$ is the SM, but also of great interest, since it
provides the upper bound on the mass of the lightest Higgs
(which is interesting for phenomenological purposes, e.g. at
LEP2). In this case the threshold correction at $M_S$ for the SM
quartic coupling $\lambda$ is:
\be
\label{threshold}
\Delta_{\rm th}\lambda=\frac{3}{16\pi^2}h_t^4
\frac{X_t^2}{M_S^2}\left(2-\frac{1}{6}\frac{X_t^2}{M_S^2}\right)
\ee
where $h_t$ is the SM top Yukawa coupling and 
$X_t=(A_t-\mu/\tan\beta)$ is the mixing in the stop mass
matrix, the parameters $A_t$ and $\mu$ being the trilinear
soft-breaking coupling in the stop sector and the supersymmetric
Higgs mixing mass, respectively. The maximum of
(\ref{threshold}) corresponds to $X_t^2=6 M_S^2$ which provides
the maximum value of the lightest Higgs mass: this case will be
referred to as the case of maximal mixing.

We have plotted in Fig.~\ref{fval3} the lightest Higgs pole mass 
$M_H$, where all NTLL corrections
are resummed to all-loop by the RG, 
as a function of $M_t$~\cite{CEQR}. From Fig.~\ref{fval3} we
can see that the present experimental band from CDF/D0 for the
top-quark mass requires $M_H\simlt 140$ GeV, while if we fix
$M_t=170$ GeV, the upper bound $M_H\simlt 125$ GeV
follows. It goes without saying
that these figures are extremely relevant for MSSM Higgs searches 
at LEP2.

\subsubsection{An analytical approximation}

We have seen~\cite{CEQR} that,
since radiative corrections are minimized for scales $Q\sim m_t$, 
when the LL RG improved Higgs mass expressions are 
evaluated at the top-quark mass scale, they reproduce the NTLL value
with a high level of accuracy, for any value of $\tan\beta$ and the 
stop mixing parameters~\cite{CEQW}
\be
\label{relmasas}
m_{H,LL}(Q^2\sim m_t^2)\sim m_{H,NTLL}.
\ee
Based on the above observation, we can work out a very accurate 
analytical approximation to $m_{H,NTLL}$ by just keeping two-loop
LL corrections at $Q^2=m_t^2$, i.e. corrections of order $t^2$, where
$t=\log(M_S^2/m_t^2)$.

Again the case $m_A\sim M_S$ is the simplest, and very illustrative, 
one. We have found~\cite{CEQW,HHH} that, in the absence of mixing
(the case $X_t=0$) two-loop corrections resum in the one-loop
result shifting the energy scale from $M_S$ (the tree-level scale)
to $\sqrt{M_S\; m_t}$. More explicitly,
\be
\label{resum}
m_H^2=M_Z^2 \cos^2 2\beta\left(1-\frac{3}{8\pi^2}h_t^2\; t\right)
+\frac{3}{2\pi^2 v^2}m_t^4(\sqrt{M_S m_t}) t
\ee
where $v=246.22$ GeV.

In the presence of mixing ($X_t\neq 0$), the run-and-match procedure
yields an extra piece in the SM effective potential 
$\Delta V_{\rm th}[\phi(M_S)]$ whose second derivative gives an
extra contribution to the Higgs mass, as
\be
\label{Deltathm}
\Delta_{\rm th}m_H^2=\frac{\partial^2}{\partial\phi^2(t)}
\Delta V_{\rm th}[\phi(M_S)]=
\frac{1}{\xi^2(t)}
\frac{\partial^2}{\partial\phi^2(t)}
\Delta V_{\rm th}[\phi(M_S)]
\ee
which, in our case, reduces to 
\be
\label{masthreshold}
\Delta_{\rm th}m_H^2=
\frac{3}{4\pi^2}\frac{m_t^4(M_S)}{v^2(m_t)}
\frac{X_t^2}{M_S^2}\left(2-\frac{1}{6}\frac{X_t^2}{M_S^2}\right)
\ee

We have compared our analytical approximation~\cite{CEQW} 
with the numerical NTLL result~\cite{CEQR} and found a difference
$\simlt 2$ GeV for all values of supersymmetric parameters.

\begin{figure}[ht]
\centerline{
\psfig{figure=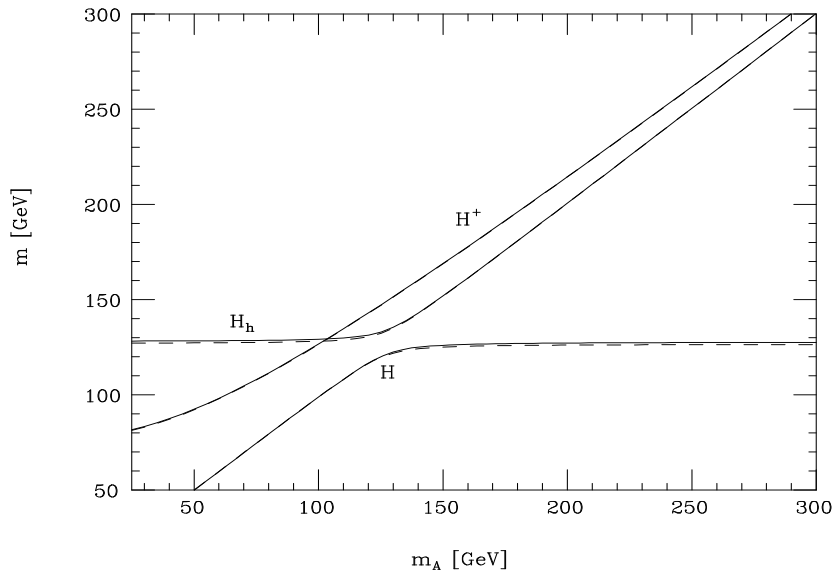,height=9.5cm,width=14cm,angle=90} }
\caption[0]
{The neutral ($H,{\cal H}\equiv H_h$ in the figure) 
and charged ($H^+$) Higgs mass spectrum 
as a function of the CP-odd Higgs mass $m_A$ for
a physical top-quark mass $M_t =$ 175 GeV and $M_S$ = 1 TeV, as
obtained from the one-loop improved RG evolution 
(solid lines) and the analytical formulae (dashed lines).
All sets of curves correspond to
$\tan \beta=$ 15 and large squark mixing, $X_t^2 = 6 M_S^2$
($\mu=0$).} 
\label{fval4}
\end{figure}

The case $m_A<M_S$ is a bit more complicated since the effective theory
below the supersymmetric scale $M_S$ is the 2HDM. However since radiative
corrections in the 2HDM are equally dominated by the top-quark, we can
compute analytical expressions based upon the LL approximation
at the scale $Q^2\sim m_t^2$. 
Our approximation~\cite{CEQW} differs from
the LL all-loop numerical resummation by $\simlt 3$ GeV, which we
consider the uncertainty inherent in the theoretical calculation,
provided the mixing is moderate and, in particular, bounded by the
condition,
\be
\label{condicion}
\left|\frac{m^2_{\;\widetilde{t}_1}-m^2_{\;\widetilde{t}_2}}
{m^2_{\;\widetilde{t}_1}+m^2_{\;\widetilde{t}_2}}\right|\simlt 0.5
\ee
where $\widetilde{t}_{1,2}$ are the two stop mass eigenstates.
In Fig.~\ref{fval4} the Higgs mass spectrum is plotted versus $m_A$.

\subsubsection{Threshold effects}

There are two possible caveats in the 
analytical approximation we have just 
presented: {\bf i)} Our expansion parameter $\log(M_S^2/m_t^2)$
does not behave properly in the supersymmetric limit $M_S\rightarrow 0$,
where we should recover the tree-level result. {\bf ii)} We have expanded
the threshold function $\Delta V_{\rm th}[\phi(M_S)]$ to order $X_t^4$.
In fact keeping the whole threshold function $\Delta V_{\rm th}[\phi(M_S)]$
we would be able to go to larger values of $X_t$ and to evaluate the
accuracy of the approximation (\ref{threshold}) and (\ref{masthreshold}).
Only then we will be able to
check the reliability of the maximum value of the
lightest Higgs mass (which corresponds to the maximal mixing) as provided
in the previous sections.
This procedure has been properly followed~\cite{CEQW,CQW} for
the most general case $m_Q\neq m_U\neq m_D$.
We have proved that keeping the exact threshold function
$\Delta V_{\rm th}[\phi(M_S)]$, and properly running its value from the
high scale to $m_t$ with the corresponding anomalous dimensions as in
(\ref{Deltathm}), produces two effects: {\bf i)} It makes a resummation
from $M_S^2$ to $M_S^2+m_t^2$ and generates as (physical) expansion
parameter $\log[(M_S^2+m_t^2)/m_t^2]$. {\bf ii)} It generates a whole
threshold function $X_t^{\rm eff}$ such that (\ref{masthreshold})
becomes
\be
\label{masthreshold2}
\Delta_{\rm th}m_H^2=
\frac{3}{4\pi^2}\frac{m_t^4[M_S^2+m_t^2]}{v^2(m_t)}
X_t^{\rm eff}
\ee
and 
\be
\label{desarrollo}
X_t^{\rm eff}=\frac{X_t^2}{M_S^2+m_t^2}
\left(2-\frac{1}{6}\frac{X_t^2}{M_S^2+m_t^2}\right)+\cdots
\ee
The numerical calculation shows~\cite{CQW} that $X_t^{\rm eff}$ 
has the maximum very close to $X_t^2=6(M_S^2+m_t^2)$,
what justifies
the reliability of previous upper bounds on the lightest Higgs mass.

\subsection{\bf The case of obese supersymmetry}

We will conclude this lecture with a very interesting case, 
where the Higgs sector of the
MSSM plays a key role in the detection of supersymmetry. 
It is the case where all supersymmetric particles are superheavy
\be
M_S \sim 1-10\ {\rm TeV}
\ee
and escape detection at LHC. 

In the Higgs sector ${\cal H},A,H^\pm$
decouple, while the $H$ couplings go the SM $\phi$ couplings 
\be
HXY\longrightarrow (\phi XY)_{\rm SM}
\ee
as $\sin(\beta-\alpha)\rightarrow 1$, or are indistinguisable
from the SM ones
\begin{eqnarray}
h_u\sin\beta & \equiv & h_u^{\rm SM} \nonumber \\
h_{d,\ell}\cos\beta & \equiv & h_{d,\ell}^{\rm SM} 
\end{eqnarray}
In this way the $\tan\beta$ dependence of the couplings, either
disappears or is absorbed in the SM couplings.

\begin{figure}[htb]
\centerline{
\psfig{figure=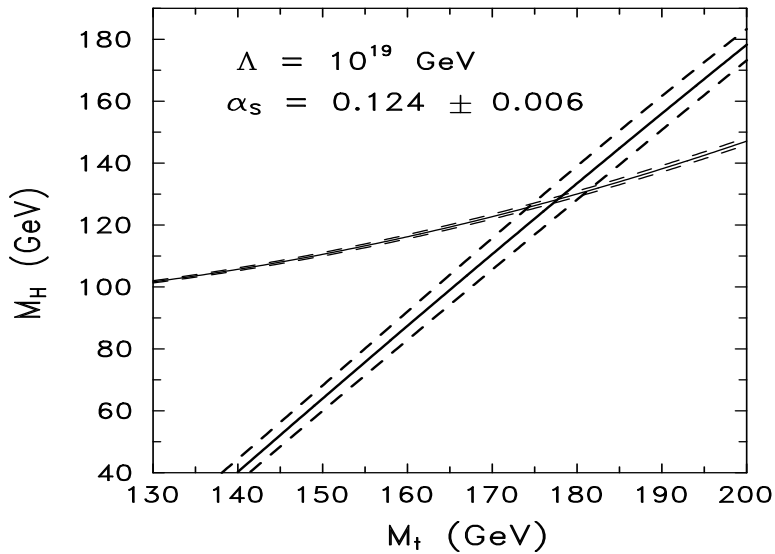,height=7.5cm,width=7cm,bbllx=5.cm,bblly=2.cm,bburx=14.5cm,bbury=15cm}}
\caption{SM lower bounds on $M_H$ (thick lines) as a function of
$M_t$, for $\Lambda=10^{19}$ GeV, from metastability requirements,
and upper bound on the lightest Higgs boson mass in the MSSM
(thin lines) for $M_S=1$ TeV.}
\label{fval5}
\end{figure}

However, from the previous sections it should be clear 
that the Higgs and top mass measurements
could serve to discriminate between the SM and its extensions, 
and to provide information about the
scale of new physics $\Lambda$. 
In Fig.~\ref{fval5}
we give the SM lower bounds on
$M_H$ for $\Lambda\simgt 10^{15}$ GeV (thick lines) and
the upper bound on the mass of the
lightest Higgs boson in the MSSM (thin lines) 
for $M_S\sim 1$ TeV. Taking, 
for instance,  $M_t=180$ GeV, close to the
central value recently reported by CDF+D0~\cite{top}, 
and $M_H\simgt 130$ GeV, the SM is
allowed and the MSSM is excluded. On the other hand, 
if $M_H\simlt 130$ GeV, then the MSSM is
allowed while the SM is excluded. Likewise there 
are regions where the SM is excluded, others
where the MSSM is excluded and others where both are permitted or 
both are excluded.

\section{Conclusion}

To conclude, we can say that the search of the Higgs boson at present
and future colliders is, not only an experimental challenge, being the
Higgs boson the last missing ingredient of the Standard Model, but
also a theoretically  appealing question from the 
more fundamental point of view of
physics beyond the Standard Model. In fact, if we are lucky enough and
the Higgs boson is detected soon (preferably at LEP2) and {\it light}, 
its detection
might give sensible information about the possible existence of
new physics. In that case, the
experimental search of the new physics should be urgent and compelling, 
since the existence of new phenomena
might be necessary for our present understanding of the physics
for energies at reach in the planned accelerators.

\section*{Acknowledgments}

Work supported in part by 
the European Union (contract CHRX-CT92-0004) and
CICYT of Spain (contract AEN95-0195).
I wish to thank my collaborators in the subjects whose results
are reported in the present lectures: M.~Carena, J.A.~Casas,
J.R.~Espinosa, A.~Riotto, C.~Wagner and F.~Zwirner. I also want
to thank A.~Riotto for his help in drawing some of the diagrams
contained in this paper.

\end{document}